\documentclass[10pt]{article}  
\usepackage{amsfonts}
\usepackage{amsmath}
\usepackage{amssymb} 
\usepackage{amsxtra}
\usepackage{pstricks,pst-plot} 
\usepackage[normalem]{ulem} 
\usepackage{rotating} 

\newtheorem{theorem}{Theorem} 
\newtheorem{lemma}{Lemma} 
\newtheorem{corollary}{Corollary} 
\newtheorem{definition}{Definition}

\newenvironment{proof}[1][Proof]{
\begin{trivlist} 
\item[\hskip \labelsep {\bfseries #1}]}
{\end{trivlist}}

\newenvironment{example}[1][Example]{\begin{trivlist}
\item[\hskip \labelsep {\bfseries #1}]}{\end{trivlist}}

\accentedsymbol{\hathatw}{\Hat{\hat w}}

\newcommand{\qed}{\nobreak \ifvmode \relax \else
      \ifdim\lastskip<1.5em \hskip-\lastskip
      \hskip1.5em plus0em minus0.5em \fi \nobreak
      \vrule height0.75em width0.5em depth0.25em\fi}

\def\Nset{\mathbb{N}} 
\def\simulates{\mapsto} 

\newcommand{\arbApp}[0]{{m}} %
\newcommand{\dczero}[0]{{0}}
\newcommand{\dcone}[0]{{1}}
\newcommand{\dctwo}[0]{{2}}
\newcommand{\dcthree}[0]{{3}}
\newcommand{\dcfour}[0]{{4}}
\newcommand{\dcfive}[0]{{5}}
\newcommand{\dcsix}[0]{{6}}
\newcommand{\dcseven}[0]{{7}}
\newcommand{\dcend}[0]{{8}}

\def \dwxout {\bgroup \markoverwith{/} \ULon}

\newcommand{\os}[3]
{\raisebox{0pt}[\height][0.3\height]{\ensuremath{\underset{
 {\makebox[0.1cm][l]{\raisebox{0.3ex}[\height]{\ensuremath{\scriptscriptstyle #3}}}}}{{#1}}{ \raisebox{-0.35ex}{\ensuremath{\scriptstyle{#2}{\hspace{0.1ex}}}}}
}}} 

\newcommand{\osa}[4]
{\raisebox{0pt}[\height][0.3\height]{\ensuremath{\underset{
 {\makebox[0.1cm][l]{\raisebox{0.3ex}[\height]{\ensuremath{
 \underset{\scriptscriptstyle #4}{\scriptscriptstyle #3} 
 }}}}}{{#1}}{ \raisebox{-0.35ex}{\ensuremath{\scriptstyle{#2}{\hspace{0.1ex}}}}}
}}}

\newcommand{\es}[3]
{\raisebox{0pt}[\height][-0.3\height]{\ensuremath{\underset{
 {\makebox[0.1cm][l]{\raisebox{0.3ex}[\height]{\ensuremath{\scriptscriptstyle #3}}}}}{{\dot #1}}{ \raisebox{-0.35ex}{\ensuremath{{\scriptstyle #2}{\hspace{0.1ex}}}}}
}}}

\newcommand{\esa}[4]
{\raisebox{0pt}[\height][-0.3\height]{\ensuremath{\underset{
 {\makebox[0.1cm][l]{\raisebox{0.3ex}[\height]{\ensuremath{
 \underset{\scriptscriptstyle #4}{\scriptscriptstyle #3} 
 }}}}}{\dot #1}{ \raisebox{-0.35ex}{\ensuremath{{\scriptstyle #2}{\hspace{0.1ex}}}}}
}}}

\newcommand{\thirdsymbol}[3]
{\raisebox{0pt}[\height][-0.3\height]{\ensuremath{\underset{
 {\makebox[0.1cm][l]{\raisebox{0.3ex}[\height]{\ensuremath{\scriptscriptstyle #3}}}}}{{\ddot #1}}{ \raisebox{-0.35ex}{\ensuremath{{\scriptstyle #2}{\hspace{0.1ex}}}}}
}}}

\newcommand{\osm}[3]
{\raisebox{0pt}[\height][\height]{\ensuremath{\underset{
 {\makebox[0.1cm][l]{\raisebox{1.93ex}[\height]{\ensuremath{\scriptscriptstyle #3}}}}}{{\dwxout{\ensuremath{#1}}}}{ \raisebox{-0.35ex}{\ensuremath{\scriptstyle{#2}{\hspace{0.1ex}}}}}
}}} %

\newcommand{\esm}[3]
{\raisebox{0pt}[\height][-0.3\height]{\ensuremath %
{\underset{
 {\makebox[0.1cm][l]{\raisebox{1.93ex}[\height]{\ensuremath{\scriptscriptstyle #3}}}}}{{\dot {\dwxout{\ensuremath{#1}}}}}{ \raisebox{-0.35ex}{\ensuremath{{\scriptstyle #2}{\hspace{0.1ex}}}}}
}}}

\newcommand{\oes}[3]{\os #1#2#3 \es #1#2#3} %
\newcommand{\oesm}[3]{\osm #1#2#3 \esm #1#2#3} %
\newcommand{\threesymb}[3]{\os #1#2#3 \es #1#2#3 \thirdsymbol #1#2#3 }
\newcommand{\osr}[3]{\os {\bar{#1}} #2 #3} %
\newcommand{\esr}[3]{\os {\Dot{\bar{#1}}} #2 #3} %
\newcommand{\thirdsr}[3]{\os {\ddot{\bar{#1}}} #2 #3} %
\newcommand{\oesr}[3]{\osr #1#2#3 \esr #1#2#3} %
\newcommand{\dwrule}[2]{\ensuremath{#1 \rightarrow #2}} %

\newcommand{\osra}[4]{\osa {\bar{#1}} #2 #3 #4} %
\newcommand{\esra}[4]{\osa {\Dot{\bar{#1}}} #2 #3 #4}

\begin{document}

\title{On the time complexity of 2-tag systems and small universal Turing machines\thanks{Proc. 47th Annual IEEE Symposium on Foundations of Computer Science (FOCS) 2006.}}
\author{Damien Woods
\\
Boole Centre for Research in Informatics\\
Department of Mathematics\\  
University College Cork, Ireland.\\\texttt{http://www.bcri.ucc.ie/\raisebox{-0.8ex}{\~}dw5}
\and
Turlough Neary
\\
TASS, Department of Computer Science\\ 
National University of Ireland, Maynooth, Ireland.\\ 
\texttt{http://www.cs.nuim.ie/\raisebox{-0.8ex}{\~}tneary}\\
}
\date{December, 2006}
\maketitle

\begin{abstract}
We show that 2-tag systems efficiently simulate Turing machines. As a corollary we find that the small universal Turing machines of Rogozhin, Minsky and others simulate Turing machines in polynomial time. This is an exponential improvement on the previously known simulation time overhead and improves a forty year old result in the area of small universal Turing machines.  
\end{abstract}

\section{Introduction}\label{sec:Intro}
It has been an open question for forty years as to whether the smallest known universal Turing machines (UTMs) are efficient simulators of Turing machines. This question is intimately related to a problem regarding the computational complexity of 2-tag systems.  

Shannon~\cite{Sha1956p} was the first to consider the question of finding the smallest possible UTM, where size is the number of states and symbols. Early attempts~\cite{Ike1958x,Wat1961p} gave small UTMs that efficiently (in polynomial time) simulate Turing machines.

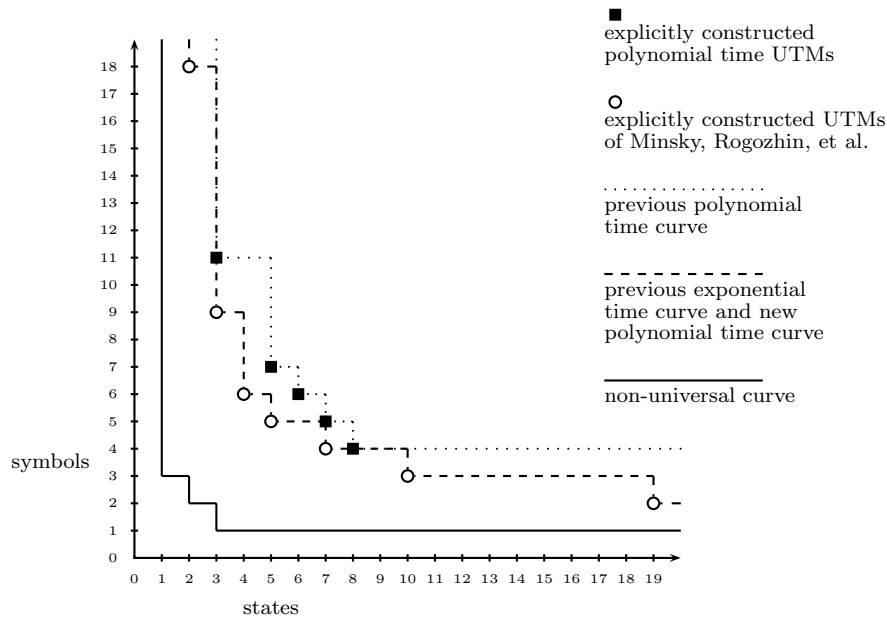
\begin{figure}
\begin{center}
\newpsobject{showgrid}{psgrid}{linestyle=dotted,subgriddiv=1,griddots=1,gridlabels=6pt}
\psset{unit=2.4ex}

\begin{pspicture}(0,-3)(22,20) 
\psset{dotsize=5pt}
\psset{dotstyle=square*}
\psdot (17.6,19.9)

\put (17.2,19.0) {\footnotesize explicitly constructed} 
\put (17.2,18.2) {\footnotesize polynomial time UTMs}

\psset{linestyle=solid}
\psset{linecolor=white}
\qdisk(17.6,16.7){0.25}
\psset{linecolor=black}
\pscircle(17.6,16.7){0.25}
\put (17.2,15.8) {\footnotesize  explicitly constructed UTMs} 
\put (17.2,15.0) {\footnotesize  of Minsky, Rogozhin, et al.}

\psset{linestyle=dotted}
\psline (17.2,13.5)(23,13.5) 
\put (17.2,12.7) {\footnotesize  previous polynomial}
\put (17.2,11.9) {\footnotesize  time curve}

\psset{linestyle=dashed, dash=3pt 3pt}
\psline (17.2,10.4)(23,10.4) 
\put (17.2,9.6) {\footnotesize  previous exponential}
\put (17.2,8.8) {\footnotesize  time curve and new}
\put (17.2,8.0) {\footnotesize  polynomial time curve}

\psset{linestyle=solid}
\psline (17.2,6.5)(23,6.5) 
\put (17.2,5.7) {\footnotesize  non-universal curve}

\psset{linestyle=solid}

{\tiny \psaxes[ticksize=1pt]{->}(20,19) }
\rput (5,-1.8) {\footnotesize states}
\put (-4.5,3.3) {\footnotesize symbols}

\psline (3,1)(20,1)
\psline (1,3)(1,19)
\psline (2,2)(3,2)
\psline (2,2)(2,3)
\psline (1,3)(2,3)
\psline (3,1)(3,2)

\psset{dotstyle=square*}

\psdot (3,11)
\psdot (5,7)
\psdot (6,6)
\psdot (7,5)
\psdot (8,4)
\psset{linestyle=dotted}
\psline (3,11)(3,19) %
\psline (3,11)(5,11)
\psline (5,7)(5,11)
\psline (5,7)(6,7)
\psline (6,6)(6,7)
\psline (6,6)(7,6)
\psline (7,5)(7,6)
\psline (7,5)(8,5)
\psline (8,4)(8,5)
\psline (8,4)(20,4) 

\psset{linestyle=dashed, dash=3pt 3pt}
\psline (2,18)(2,19) 
\psline (2,18)(3,18)
\psline (3,9)(3,18)
\psline (3,9)(4,9)
\psline (4,6)(4,9)
\psline (4,6)(5,6)
\psline (5,5)(5,6)
\psline (5,5)(7,5)
\psline (7,4)(7,5)
\psline (7,4)(10,4)
\psline (10,3)(10,4)
\psline (10,3)(19,3)
\psline (19,2)(19,3)
\psline (19,2)(20,2) 

\psset{linestyle=solid}
\psset{linecolor=white}
\qdisk(2,18){0.25}
\psset{linecolor=black}
\pscircle(2,18){0.25}

\psset{linestyle=solid}
\psset{linecolor=white}
\qdisk(3,9){0.25}
\psset{linecolor=black}
\pscircle(3,9){0.25}

\psset{linestyle=solid}
\psset{linecolor=white}
\qdisk(4,6){0.25}
\psset{linecolor=black}
\pscircle(4,6){0.25}

\psset{linestyle=solid}
\psset{linecolor=white}
\qdisk(5,5){0.25}
\psset{linecolor=black}
\pscircle(5,5){0.25}

\psset{linestyle=solid}
\psset{linecolor=white}
\qdisk(7,4){0.25}
\psset{linecolor=black}
\pscircle(7,4){0.25}

\psset{linestyle=solid}
\psset{linecolor=white}
\qdisk(10,3){0.25}
\psset{linecolor=black}
\pscircle(10,3){0.25}

\psset{linestyle=solid}
\psset{linecolor=white}
\qdisk(19,2){0.25}
\psset{linecolor=black}
\pscircle(19,2){0.25}
\end{pspicture}
\caption{State-symbol plot of small UTMs. The plot shows the polynomial time curve induced by our previous UTMs~\cite{NW2006p--acc}, the exponential time curve of Minsky, Rogozhin and others~\cite{Bai2001c,KR2001c,Min1962p,Rog1996p}, and the non-universal %
curve for which there are no UTMs~\cite{Her1968c,Kud1996p,Pav1973p,Pav1978p}. The present paper improves the polynomial time curve so that it coincides with the previous exponential time curve. Our result shows that a polynomial time UTM exists for each state-symbol pair that is on, above, and to the right of the new polynomial time curve.}\label{fig:PECurve-april06}
\end{center}
\end{figure}

In the early 1960s Cocke and Minsky~\cite{CM1964p} showed that 2-tag systems simulate Turing machines, but in an exponentially slow fashion. Minsky~\cite{Min1962p} found a small 7-state, 4-symbol UTM that simulates 2-tag systems in polynomial time. So this small UTM simulates Turing machines via the following sequence of simulations
\begin{equation}\label{eq:MinskyRogozhinSim}
\textrm{Turing machine}\simulates \textrm{2-tag system} \simulates \textrm{small UTM}
\end{equation}
where $A\simulates B$ denotes that $A$ is simulated by $B$. Later Rogozhin~\cite{Rog1996p} and others~\cite{Bai2001c,KR2001c,Rob1991p} used Minsky's technique of simulation via~\eqref{eq:MinskyRogozhinSim} to find small UTMs for a range of state-symbol pairs (see Figure~\ref{fig:PECurve-april06}). All of these small UTMs efficiently simulate 2-tag systems. However since the 2-tag simulation of Turing machines is exponentially slow it has remained an open problem as to whether these UTMs can be made to run in polynomial time.

For our result we replace~\eqref{eq:MinskyRogozhinSim} with the following sequence of simulations
\begin{align*}%
& \textrm{Turing machine} \simulates  \textrm{cyclic tag system}\\
& \simulates \textrm{2-tag system} \simulates \textrm{small UTM} %
\end{align*}
Specifically, in this paper we show that 2-tag systems efficiently simulate cyclic tag systems. In a recent paper~\cite{NW2006c} we have shown that cyclic tag systems efficiently simulate Turing machines. Thus the present paper provides an important piece of the puzzle as far as the computational complexity of small UTMs and 2-tag systems are concerned. Our main results states:
\begin{theorem}\label{thm:2-tagSimTM}
Given a single tape deterministic Turing machine $M$ that computes in time~$t$ then there is a 2-tag system~$T_M$ that simulates the computation of~$M$ and computes in polynomial time $O(t^{4}(\log t)^{2})$.
\end{theorem}
This immediately gives the following interesting result.
\begin{corollary}\label{cor:UTMsSimTM}
The small UTMs of Minsky, Rogozhin and others~\cite{Bai2001c,KR2001c,Min1962p,Rob1991p,Rog1996p} are polynomial time, $O(t^{8}(\log t)^{4})$, simulators of Turing machines.
\end{corollary}

Before our result it was entirely plausible that there was an exponential trade-off between UTM program size complexity, and time/space complexity; the smallest UTMs seemed to be exponentially slow. However our result shows there is currently little evidence for such a claim.

Early examples of efficient small UTMs were found by Ikeno and Watanabe~\cite{Ike1958x,Wat1961p}.  Prior to the present paper the smallest known polynomial time UTMs were to be found in~\cite{NW2006p--acc}. However these efficient machines are not as small as those of Rogozhin et al., hence the present paper represents a significant size improvement when considering small polynomial time UTMs. This improvement is illustrated in Figure~\ref{fig:PECurve-april06}.

There are numerous other applications of Theorem~\ref{thm:2-tagSimTM}. The technique of simulation via 2-tag systems is at the core of many results in the broad survey by Margenstern~\cite{Mar2000p}; our result exponentially improves the time overheads in many~\cite{LN1990p,Mar1993c,Mar1995c,Mar1995bc,Rob1971p} of these constructions. 
Another example of the use of small efficient UTMs is given by  Levin and Venkatesan~\cite{LV2001m,VL1988c} who used the small 8-state, 5-symbol, polynomial time UTM of Watanabe's~\cite{Wat1961p} to show the average case NP-completeness of a graph colouring problem. %

In the present paper, the phrase ``small UTMs'' refers to Turing machines that obey the standard definitions. Recently Cook~\cite{Coo2004p} has found universal machines that are significantly smaller than those discussed in the present paper. Cook's machines simulate the cellular automaton Rule 110, which Cook showed to be universal via an impressive simulation. However these machines are generalisations of standard Turing machines: their blank tape consists of an infinitely repeated word to the left and another to the right. Intuitively, this change of definition makes quite a difference, especially since Cook encodes a (possibly universal) program in one of these repeated words. It should be noted that Cook's UTMs are exponentially slow; in a recent paper~\cite{NW2006c} we have improved their simulation time overhead to polynomial.

\subsection{Preliminaries}
Let $\Nset=\{0,1,2,\ldots\}$, $\Sigma$ denote a finite alphabet, and~$\epsilon$ denote the empty word. The function $\Vert\,\Vert : \Sigma^{\ast}\times\Sigma^{\ast} \rightarrow \Nset$ is written as $\Vert w_1,w_2 \Vert$ and gives the number of occurrences of the word~$w_2$ in~$w_1$. For example $\Vert 1101001,01 \Vert =2$. All logs are in base~2. Throughout the paper~$a$ and~$\#$ are constant symbols, and $x,x_i \in\{0,1\}$. 
We define the {\em parity} of a word to be {\em odd} if the read (leftmost) symbol is undotted and {\em even} if the read symbol is dotted.

\section{2-tag systems}\label{sec:2tagDef}
Tag systems where introducted by Post~\cite{Pos1943p}, and 2-tag systems were shown to be universal by Cocke and Minsky~\cite{CM1964p}.
\begin{definition}[2-tag system]\label{def:2TS}
A tag system consists of a finite alphabet of symbols $\Sigma$, a finite set of rules $R : \Sigma\rightarrow\Sigma^{\ast}$ and a deletion number $\beta\in\Nset$, $\beta\geqslant 1$. For a 2-tag system $\beta = 2$.
\end{definition}
The 2-tag systems we consider are deterministic (or equivalently, monogenic~\cite{Coo1966p,Wan1963p}). The computation of a 2-tag system acts on a dataword $w=\sigma_1 \sigma_2\ldots \sigma_l$. The entire configuration is given by $w$. In a computation step, the symbols~$\sigma_1 \sigma_2$ are deleted and if there is a rule for $\sigma_1$, i.e.\ a rule of the form $\sigma_1\rightarrow \sigma_{l+1}\ldots \sigma_{l+c}$, then the word $\sigma_{l+1}\ldots \sigma_{l+c}$ is appended. We write $w_1\vdash w_2$ when the dataword (configuration) $w_2$ is obtained from $w_1$ via a single computation step:
\begin{equation*}
\sigma_1 \sigma_2 \sigma_3\ldots \sigma_l \vdash \sigma_3\ldots \sigma_l \sigma_{l+1}\ldots \sigma_{l+c}
\end{equation*}
where $\sigma_1\rightarrow \sigma_{l+1}\ldots \sigma_{l+c} \in R$.\hspace{0.2ex} A 2-tag system completes its computation if~(i)~$|w|< \beta $, or (ii)~it enters a repeating sequence of configurations, or (iii)~there is no rule for the leftmost symbol. 
The complexity measures of time and space are defined in the obvious way. Given a word~$w$, we use the term {\em round} to describe~$\lceil \vert w\vert/2\rceil$ computation steps that traverse~$w$ exactly once~\cite{Coo1966p,Wan1963p}.

We often write 2-tag symbols in pairs. The second (even numbered) symbol is dotted to distinguish it from the first. In the sequel we encode binary symbols in the following way, $1$ is encoded as $\oes{1}{{}}{{}}$ and 0 as $\oes{0}{{}}{{}}$.  Also a single pair of symbols is distinguished by being `barred': $\osr{0}{{}}{\,}\esr{0}{{}}{\,}$ or $\osr{1}{{}}{\,}\esr{1}{{}}{\,}$. 
So an example encoding of the word~$11010$ is~$\osr{1}{{}}{{}}\esr{1}{{}}{{}}\oes{1}{{}}{{}}\oes{0}{{}}{{}}\oes{1}{{}}{{}}\oes{0}{{}}{{}}$. 
\begin{lemma}\label{lem:oddEven}Let 
 $w=\osr{x}{0}{{}}\esr{x}{0}{{}}x_1x_2\ldots x_l \in\{\osr{0}{{}}{{}}\esr{0}{{}}{{}},\osr{1}{{}}{{}}\esr{1}{{}}{{}}\}\{0,1\}^{\ast}$. Then there is a 2-tag system $T$ that tests whether~$\vert  w\vert $ is odd or even in exactly~$\lfloor\vert w\vert /2\rfloor+1$ timesteps.\end{lemma}
\begin{proof}
The 2-tag system has 6 rules, $R=\{\osr{x}{{}}{{}}\rightarrow \osr{x}{{}}{1}\esr{x}{{}}{1}, x\rightarrow \epsilon,\osr{x}{{}}{1}\rightarrow\os{x}{{}}{2}\}$ where $x\in\{0,1\}$. Initially $T$ reads the leftmost symbol $\osr{x}{0}{{}}$ of $w$. After one round, if the read symbol is dotted then the output is the single symbol $\esr{x}{{}}{1}$ signifying that~$|w|$ is odd. Otherwise the output is $\os{x}{{}}{2}$ signifying that~$|w|$ is even.\qed \end{proof}

Despite its simplicity, the proof idea of Lemma~\ref{lem:oddEven} constitutes one of the main ingredients of Cocke and Minsky's (exponentially slow) 2-tag simulation of Turing machines~\cite{CM1964p}. The 2-tag system dataword encodes an arbitrary TM configuration as two unary numbers. The left side of the tape is encoded as one unary number, the right side as another. Their simulation makes use of repeated tests for oddness and evenness of dataword length. Also doubling and halving of dataword length is used to read and write to the simulated tape.

\section{Cyclic tag systems}
Cyclic tag systems were introduced by Cook~\cite{Coo2004p}.
\begin{definition}[cyclic tag system~\cite{Coo2004p}]\label{def:CyclicTS}
A cyclic tag system $C=\alpha_{0},\alpha_{1},\ldots,\alpha_{p-1}$ is a list of binary words $\alpha_{\arbApp}\in\{0,1\}^{\ast}$ called appendants.
\end{definition}
A {\em configuration} of a cyclic tag system consists of (i) a {\em marker} that points to a single appendant $\alpha_{\arbApp}$ in $C$, and (ii) a dataword $w = x_0 x_1\ldots x_l \in\{0,1\}^*$. Intuitively the list $C$ is a {\em program} with the marker pointing to instruction $\alpha_{\arbApp}$. At the initial configuration the marker points to appendant $\alpha_0$ and $w$ is the binary input word. 
\begin{definition}[computation step of a cyclic tag system]\label{def:OperationJTS} 
A computation step is deterministic and acts on a configuration in one of two ways:
\begin{itemize}
\item If $x_0 = 0$ then $x_0$ is deleted and the marker moves to appendant $\alpha_{({\arbApp}+1)\mod p}$. 
\item If $x_0 = 1$ then $x_0$ is deleted, the word $\alpha_{\arbApp}$ is appended onto the right end of $w$,  and the marker moves to appendant $\alpha_{({\arbApp}+1)\mod p}$. 
\end{itemize}
\end{definition}

A cyclic tag system completes its computation if (i) the dataword is the empty word, or (ii) it enters a repeating sequence of configurations. The complexity measures of time and space are defined in the obvious way.

\begin{example}
\textit{(cyclic tag system computation)} Let $C=00,010,11$ be a cyclic tag system with input word $011$. Below we give the first four steps of the computation. In each configuration $C$ is given on the left with the marked appendant highlighted in bold font. 
\begin{eqnarray*}
\begin{aligned}  
            & \pmb{00},010,11 \quad 011 & \vdash\quad  & 00,\pmb{010},11 \quad 11 & \\
\vdash\quad & 00,010,\pmb{11} \quad 1010 & \vdash\quad  & \pmb{00},010,11 \quad 01011 & \\
\vdash\quad & 00,\pmb{010},11 \quad 1011 & \vdash\quad  &\ldots
\end{aligned}
\end{eqnarray*}
\end{example}
We write an arbitrary single step of a cyclic tag system computation as
\begin{equation}\label{eq:cyclicTagStepToBeSim}
\begin{aligned} %
& \alpha_0  
 ,\ldots,
 \alpha_{{\arbApp}-1},
 \pmb{\alpha_{{\arbApp}}},
 \alpha_{{\arbApp}+1},\ldots,\alpha_{p-1} \quad x_0 x_1 \ldots x_l \\
\vdash\quad   &
\alpha_0,
\ldots,
\alpha_{{\arbApp}},
\pmb{\alpha_{{\arbApp}+1}},
\alpha_{{\arbApp}+2},
\ldots,
\alpha_{p-1} \quad  x_1\!\ldots x_l x_{l+1}\!\ldots x_{l+c}
\end{aligned} %
\end{equation}
where $x\in\{0,1\}$, and as usual if $x_0 = 0$ then $x_{l+1} \ldots x_{l+c}=\epsilon$, otherwise if~$x_0 = 1$ then $x_{l+1} \ldots x_{l+c} = \alpha_{\arbApp}\in\{0,1\}^{c}, c\in\Nset$.

Cook~\cite{Coo2004p} used the universality of cyclic tag systems to show that Rule~110, a binary one-dimensional cellular automaton, is universal. Recently we have improved on Cook's work by showing that cyclic tag systems simulate Turing machines in polynomial time:
\begin{theorem}[\cite{NW2006c}]\label{lem:CycTagSysSimTM}
Let $M$ be a single-tape deterministic Turing machine that computes in time~$t$. Then there is a cyclic tag system $C_M$ that simulates the computation of $M$ in time $O(t^{3} \log t)$.
\end{theorem}
In order to calculate this upper bound we substitute space bounds for time bounds whenever possible in the analysis.

\section{2-tag systems efficiently simulate cyclic tag systems}
In this section we prove Theorem~\ref{thm:2-tagSimTM}. 

\subsection{Encoding}
Cyclic tag systems use a binary alphabet and program control is determined by the read symbol and the value of the program instruction marker. On the one hand 2-tag systems seem more general than cyclic tag systems as an arbitrary constant (independent of input length) sized alphabet is permitted. On the other hand 2-tag systems seem more restricted in that program control is determined solely by the read symbol. 

Because of this restriction we use a large number of symbols in our construction. The number of such symbols is a constant that is independent of input length, but is dependent on our simulation algorithm and the size of the simulated cyclic tag system program.  In our encoding we decorate symbols with dots ($\es{x}{{}}{{}},\thirdsymbol{x}{{}}{{}}$), bars (\osr{x}{{}}{{}}) and under-indexes~$(\os{x}{{}}{j})$. These decorations are used for algorithm control flow.

\begin{definition}[2-tag input encoding]\label{def:2tagInputEnc} The cyclic tag system input dataword $w=x_0x_1\ldots x_n\in\{0,1\}^{\ast}$ is encoded as the 2-tag dataword
\begin{equation*}
\hat{w}=
\oesr{x}{0}{1}\; 
\oes{x}{1}{1}\; 
\oes{x}{2}{1}\ldots
\oes{x}{n}{1}\; 
\oes{a}{{}}{1}\; \oes{a}{{}}{1} \ldots \oes{a}{{}}{1}
\end{equation*}
where the number of $\oes{a}{{}}{\,}$ pairs in $\hat{w}$ is $\Vert\hat{w},\oes{a}{{}}{\,}\Vert=2^{\lceil\log (n+1)\rceil}$ and  the extra whitespace between symbol pairs is for human readability purposes only. 
\end{definition}

This encoding is computable in logspace. The subword $\oes{a}{{}}{\,}\,\oes{a}{{}}{\,}\ldots\oes{a}{{}}{\,}$ is used as a {\em counter} and its {\em value} $\Vert\hat{w},\oes{a}{{}}{\,}\Vert$ is used extensively in our algorithms below. 

An arbitrary (not necessarily input) cyclic dataword is encoded similarly to Definition~\ref{def:2tagInputEnc} except that the counter is `embedded' in $w$. Specifically if $w=x_0x_1\ldots x_l$ then
\begin{equation}\label{eq:2tagArbitraryWordEnc}
\hat{w}=
\oesr{x}{0}{j}\; \oes{x}{1}{j}\ldots\oes{x}{{i}}{j}\; 
\oes{a}{{}}{j}\ldots\oes{a}{{}}{j}\; 
\oes{x}{{i+1}}{j}\ldots\oes{x}{{l}}{j}
\end{equation}
for some $i\in\{0,\ldots,l\}$ and $j\in\Nset$. Furthermore the counter value is some power of 2 strictly greater than $l$, that is
\begin{equation}\label{eq:counterValue}
\Vert\hat{w},\oes{a}{{}}{\,}\Vert = 2^{\lceil\log l'\rceil}
\end{equation}
where $l'\in\Nset$, $l' > l$. 

In our simulation the counter does not grow too large. As will be shown, the counter value is never more than double the longest dataword length that occurs in the simulated cyclic tag system computation.

\subsection{The simulation}\label{sec:sim}
We wish to show that there is a 2-tag system that simulates an arbitrary single step of a cyclic tag system computation, as defined in Equation~\eqref{eq:cyclicTagStepToBeSim}. We decompose Equation~\eqref{eq:cyclicTagStepToBeSim} into three conceptual steps: (i)~if $x_0 = 1$ then simulate the rule $x_0\rightarrow \alpha_{\arbApp}$ by appending $\alpha_{\arbApp} = x_{l+1} \ldots x_{l+c}$, (ii)~set $x_1$ to be the new read symbol and delete $x_0$, (iii)~increment the program marker ${\arbApp}$ so that the next appendant is $\alpha_{({\arbApp}+1 \mod p)}$.

We begin by giving a simulation of (ii). Informally speaking, Lemma~\ref{lem:passBar} states that there is a 2-tag system that (efficiently) moves the `bar' forward by one symbol pair.  The main difficulty is in distinguishing $\oes{x}{1}{1}$ from the other unbarred symbol pairs $\oes{x}{2}{2}, \ldots, \oes{x}{l}{l}$. This Lemma is stated for the case that the counter value is the next power of two greater than the encoded dataword length.

\begin{lemma}\label{lem:passBar} 
Given a word of the form
\begin{equation*}
\hat{w}=
\oesr{x}{0}{1}\; 
\oes{x}{1}{1}\ldots\oes{x}{{i}}{1}\; 
\oes{a}{{}}{1}\ldots\oes{a}{{}}{1}\;  
\oes{x}{{i+1}}{1}\ldots\oes{x}{{l}}{1}   
\end{equation*}
 then there is a 2-tag system $T$ that computes
\begin{equation*}
\hathatw = 
\oesr{x}{1}{6}\ldots\oes{x}{{i}}{6}\;  
\oes{a}{{}}{6}  %
\ldots\oes{a}{{}}{6}\;  
\oes{x}{{i+1}}{6} %
\ldots\oes{x}{{l}}{6} 
\end{equation*}
in time $O(l \log l)$. Here $\Vert\hat{w},\oes{a}{{}}{\,}\Vert= \Vert\hathatw ,\oes{a}{{}}{\,}\Vert =2^{\lceil\log (l+1)\rceil}$ and $i \in\{0,\ldots,l\}$.
\end{lemma}
\begin{proof}[Proof idea] There are 5 stages to the 2-tag algorithm. We let $\hat{w}_0 = \hat{w}$ and let~$\hat{w}_k$ denote the output of the $k^{\textrm{th}}$ iteration of the 5 stages. 

In Stages 1 to 3 of iteration $k$ we compute $\lceil\Vert \hat{w}_{k-1},\oes{x}{{}}{{}} \Vert /2 \rceil$, by marking every second pair of $\oes{x}{{}}{{}}$ symbols (we mark the even numbered pairs). Then in Stages 4 and 5 we halve $\Vert \hat{w}_{k-1},\oes{a}{{}}{{}} \Vert$ by marking every second pair of~$\oes{a}{{}}{{}}$ symbols (again we mark the even numbered pairs). We then return to Stage 1 and iterate until $\Vert \hat{w}_k,\oes{a}{{}}{{}} \Vert=1$: the counter has an odd value for the first time and we detect this. The number of fully completed iterations, and final value for~$k$, is~$k=\log \Vert \hat{w},\oes{a}{{}}{{}} \Vert$. At this point~$\oes{x}{1}{{}}$ is the only pair of unmarked~$\oes{x}{{}}{{}}$ symbols in~$\hat{w}$, and so~$\oes{x}{1}{{}}$ is isolated (unique) from the other symbol pairs in~$\hat{w}$.
We delete~$\osr{x}{0}{{}}\esr{x}{0}{{}}$. The uniqueness of $\oes{x}{1}{{}}$ enables the rule $\os{x}{1}{{}}\rightarrow\osr{x}{1}{{}}\esr{x}{1}{{}}$ to be executed successfully. 
\end{proof}
\begin{proof}[Proof details]
As usual let $x\in\{0,1\}$. Here we specify 2-tag rules, and take the reader through a single (the first) iteration of these rules, for illustration purposes we choose~$i=l$.

In Stage 1 we begin with a word of the form
\begin{equation*}
\oesr{x}{0}{1}\; 
\oes{x}{1}{1}\; 
\oes{x}{2}{1}
\ldots
\oes{x}{l}{1}\; 
\oes{a}{{}}{1}\; \oes{a}{{}}{1} \ldots \oes{a}{{}}{1}
\end{equation*}
Stage 1 consists of the rules: 
\begin{equation*}%
\{\dwrule{\osr{x}{{}}{1}}{\osr{x}{{}}{2}\esr{x}{{}}{2}},
\dwrule{\os{x}{{}}{1}}{\threesymb{x}{{}}{2}},
\dwrule{\osm{x}{{}}{1}}{\oesm{x}{{}}{2}},
\dwrule{\os{a}{{}}{1}}{\oes{a}{{}}{2}},
\dwrule{\osm{a}{{}}{1}}{\oesm{a}{{}}{2}}\}
\end{equation*}
as well as a few more rules that are given below. After one round we have 
\begin{equation}\label{eq:PassBarSimStage1}
\oesr{x}{0}{2}\; 
\threesymb{x}{1}{2}\;
\threesymb{x}{2}{2}%
\ldots
\threesymb{x}{l}{2}\; 
\oes{a}{{}}{2}\; \oes{a}{{}}{2} \ldots \oes{a}{{}}{2}
\end{equation}
The Stage 2 rules are
\begin{align*}%
\{ 
& \dwrule{\osr{x}{{}}{2}}{\osr{x}{{}}{3}\esr{x}{{}}{3}},
\dwrule{\os{x}{{}}{2}}{\oes{x}{{}}{3}},
\dwrule{\es{x}{{}}{2}}{\oesm{x}{{}}{3}},
\dwrule{\thirdsymbol{x}{{}}{2}}{\epsilon},
\dwrule{\osm{x}{{}}{2}}{\oesm{x}{{}}{3}},
\dwrule{\esm{x}{{}}{2}}{\oesm{x}{{}}{3}},\\
& \dwrule{\os{a}{{}}{2}}{\oes{a}{{}}{3}},
\dwrule{\es{a}{{}}{2}}{\oes{a}{{}}{3}},
\dwrule{\osm{a}{{}}{2}}{\oesm{a}{{}}{3}},
\dwrule{\esm{a}{{}}{2}}{\oesm{a}{{}}{3}}
\}
\end{align*}
Continuing from~\eqref{eq:PassBarSimStage1}, after one round we see that every second (even numbered) pair of $\oes{x}{{}}{{}}$ is marked
\begin{equation*}%
\oesr{x}{0}{3}\; 
\oes{x}{1}{3}\;
\oesm{x}{2}{3}\;  
\oes{x}{3}{3}\;
\oesm{x}{4}{3}%
\ldots
\oes{x}{{l-1}}{3}\; 
\oesm{x}{l}{3}\; 
\oes{a}{{}}{3}\; 
\oes{a}{{}}{3} 
\ldots 
\oes{a}{{}}{3}
\end{equation*}
where (for illustration purposes only) we assume that $l$ is even. 

The Stage 3 rules are:
\begin{align*}%
\{ 
& \dwrule{\osr{x}{{}}{3}}{\osr{x}{{}}{4}\esr{x}{{}}{4}},
\dwrule{\os{x}{{}}{3}}{\oes{x}{{}}{4}},
\dwrule{\osm{x}{{}}{3}}{\oesm{x}{{}}{4}}, 
\dwrule{\os{a}{{}}{3}}{\oes{a}{{}}{4}},
\dwrule{\osm{a}{{}}{3}}{\oesm{a}{{}}{4}}, \\ 
& \dwrule{\esr{x}{{}}{3}}{\#\osr{x}{{}}{{4}}\esr{x}{{}}{{4}}},
\dwrule{\es{x}{{}}{3}}{\oes{x}{{}}{4}},
\dwrule{\esm{x}{{}}{3}}{\oesm{x}{{}}{4}},
\dwrule{\es{a}{{}}{3}}{\oes{a}{{}}{4}},
\dwrule{\esm{a}{{}}{3}}{\oesm{a}{{}}{4}}
\}
\end{align*}
We enter Stage 3 by reading either a dotted symbol (there was an odd number of unmarked $\oes{x}{{}}{{}}$ pairs in Stage 1) or undotted  symbol (there was an even number of unmarked $\oes{x}{{}}{{}}$ pairs in Stage 1). Stage 3 begins by checking this (in one step); if the parity is even, that is the 2-tag system is reading dotted symbols, then a~$\#$ symbol is appended to restore the parity to odd after one round.
On completion of Stage 3 we are reading an undotted symbol:
\begin{equation*}%
\oesr{x}{0}{4}\; 
\oes{x}{1}{4}\;
\oesm{x}{2}{4}\;  
\oes{x}{3}{4}\;
\oesm{x}{4}{4}  
\ldots
\oes{x}{{l-1}}{4}\; 
\oesm{x}{l}{4}\; 
\oes{a}{{}}{4}\; \oes{a}{{}}{4} \ldots \oes{a}{{}}{4}
\end{equation*}

In Stages 4 and 5 of iteration $k$ we halve the value of the counter (we compute $\Vert\hat{w}_{k},\oes{a}{{}}{\,}\Vert = \Vert\hat{w}_{k-1},\oes{a}{{}}{\,}\Vert /2$), in a similar fashion to Stages 1 to 3. The Stage 4 rules are
\begin{equation*}%
\{ 
 \dwrule{\osr{x}{{}}{4}}{\osr{x}{{}}{5}\esr{x}{{}}{5}},
\dwrule{\os{x}{{}}{4}}{\oes{x}{{}}{5}},
\dwrule{\osm{x}{{}}{4}}{\oesm{x}{{}}{5}},  
 \dwrule{\os{a}{{}}{4}}{\threesymb{a}{{}}{5}},
\dwrule{\osm{a}{{}}{4}}{\oesm{a}{{}}{5}}
\}
\end{equation*}
Which after one round gives
\begin{equation*}%
\oesr{x}{0}{5}\; 
\oes{x}{1}{5}\;
\oesm{x}{2}{5}  
\ldots
\oesm{x}{l}{5}\; 
\threesymb{a}{{}}{5}\; \threesymb{a}{{}}{5} \ldots \threesymb{a}{{}}{5}
\end{equation*}
The Stage 5 rules then halve the counter value:
\begin{align*}%
\{ 
& \dwrule{\osr{x}{{}}{5}}{\osr{x}{{}}{1}\esr{x}{{}}{1}},
\dwrule{\os{x}{{}}{5}}{\oes{x}{{}}{1}},
\dwrule{\osm{x}{{}}{5}}{\oesm{x}{{}}{1}},  
\dwrule{\es{x}{{}}{5}}{\oes{x}{{}}{1}},
\dwrule{\esm{x}{{}}{5}}{\oesm{x}{{}}{1}},
\\ 
& \dwrule{\os{a}{{}}{5}}{\oes{a}{{}}{1}}, 
\dwrule{\es{a}{{}}{5}}{\oesm{a}{{}}{1}},
\dwrule{\thirdsymbol{a}{{}}{5}}{\epsilon},
\dwrule{\osm{a}{{}}{5}}{\oesm{a}{{}}{1}},
\dwrule{\esm{a}{{}}{5}}{\oesm{a}{{}}{1}}\}
\end{align*}
Continuing our computation we get:
\begin{equation*}%
\oesr{x}{0}{1}\; 
\oes{x}{1}{1}\;
\oesm{x}{2}{1}
\ldots
\oesm{x}{l}{1}\; 
\oes{a}{{}}{1}\; 
\oesm{a}{{}}{1}\; 
\oes{a}{{}}{1}\; 
\oesm{a}{{}}{1} %
\ldots 
\oes{a}{{}}{1}\;
\oesm{a}{{}}{1}
\end{equation*}
which switches control back to Stage 1.

Each iteration of the 5 stages halves the counter value. After $\log \Vert \hat{w},\oes{a}{{}}{{}} \Vert$ iterations the counter has value 1, this causes the output from Stage 4 to be of odd length for the first time. This in turn switches parity to even (dotted symbols) during Stage 5, which is detected at the beginning of Stage 1, by the rules:
\begin{equation*}%
\{
\dwrule{\esr{x}{{}}{1}}{\#},
\dwrule{\es{x}{{}}{1}}{\osr{x}{{}}{6}\esr{x}{{}}{6}},
\dwrule{\esm{x}{{}}{1}}{\oes{x}{{}}{6}},
\dwrule{\es{a}{{}}{1}}{\oes{a}{{}}{6}},
\dwrule{\esm{a}{{}}{1}}{\oes{a}{{}}{6}}
\}
\end{equation*}
The first of these rules deletes $x_0$ in one step. The second of these rules passes the bar forward by one symbol pair, while~2 of the others unmark the remaining symbols.
\begin{equation*}%
\oesr{x}{1}{6}\; 
\oes{x}{2}{6}\;
\oes{x}{3}{6}  
\ldots
\oes{x}{l}{6}\; 
\oes{a}{{}}{6}\; 
\oes{a}{{}}{6}
\ldots 
\oes{a}{{}}{6}
\end{equation*}
The $\#$ symbol restores the parity to odd so that we read undotted symbols (in subsequent computations in this paper). 

For our example iteration we chose $i=l$. If the dataword $\hat{w}$ is in the more general form $i \in\{0,\ldots,l\}$ given by the lemma statement then the same proof holds; our rules are such that embedding the counter inside the dataword does not affect parity in a way that would change the algorithm control flow.
\qed\end{proof}

The following lemma provides much of the mechanics required for simulation of the appending of a cyclic tag system appendant (point (i) from the introductory paragraph of Section~\ref{sec:sim}). Simulating the appending is straightforward, the main work is in maintaining the equality in Equation~\eqref{eq:counterValue}. Thus, after appending, if the simulated dataword has length that is at least the counter value then we double the counter to satisfy Equation~\eqref{eq:counterValue}. The Lemma is stated for the case that the counter value is initially the next power of two greater than the encoded dataword length.

\begin{lemma}\label{lem:simAppend}
Given a word of the form
\begin{equation*}%
\hat{w}= 
\osr{1}{{}}{\dczero}
\esr{1}{{}}{\dczero}\;
\oes{x}{1}{\dczero}\ldots\oes{x}{{i}}{\dczero}\; 
\oes{a}{{}}{\dczero} \; 
\oes{a}{{}}{\dczero}
\ldots
\oes{a}{{}}{\dczero}\; 
\oes{x}{{i+1}}{\dczero} %
\ldots\oes{x}{{l}}{\dczero}
\end{equation*}
where $\Vert\hat{w},\oes{a}{{}}{\,}\Vert=2^{\lceil\log (l+1)\rceil}$ and $i\in\{0,\ldots,l\}$, then there is a 2-tag system~$T$ that computes 
\begin{equation*}%
\begin{aligned}
\hathatw = & \
\osr{1}{{}}{\dcend}
\esr{1}{{}}{\dcend}\;
\oes{x}{1}{\dcend}\ldots\oes{x}{{i}}{\dcend}\; 
\oes{a}{{}}{\dcend}\;
\oes{a}{{}}{\dcend} 
\ldots
\oes{a}{{}}{\dcend}\; 
\oes{x}{{i+1}}{\dcend}
\ldots
\oes{x}{{l}}{\dcend}\; %
\oes{x}{{l+1}}{\dcend}
\ldots\oes{x}{{l+c}}{\dcend}
\end{aligned}
\end{equation*} 
where~$c \leqslant 2^{\lceil\log (l+1)\rceil}$, $\Vert\hathatw,\oes{a}{{}}{\,}\Vert=2^{\lceil\log (l+c+1)\rceil}$. $T$ completes this computation in time~$O(l \log l)$.
\end{lemma}
\begin{proof}
By applying the rule 
$$\{
\dwrule{\osr{1}{{}}{\dczero}}{\oes{x}{{l+1}}{\dczero} \ldots\oes{x}{{l+c}}{\dczero}\,\osr{1}{{}}{\dcone}\esr{1}{{}}{\dcone}}
\}$$ to $\hat{w}$ we get a word denoted $\hat{w}_0$ that is of a similar form to $\hathatw$ except that $\Vert\hat{w_0},\oes{a}{{}}{\,}\Vert=2^{\lceil\log (l+1)\rceil}$, i.e.\ the counter has not yet been updated to the correct value of $\Vert\hathatw,\oes{a}{{}}{\,}\Vert = 2^{\lceil\log (l+c+1)\rceil}$. The remainder of the proof is concerned with updating the counter.

We let~$\hat{w}_k$ denote the output of the $k^{\textrm{th}}$ iteration of the 4 stages. The rules for Stages $\dcone$ to $\dcfour$ are of a similar flavour to those used in the proof of Lemma~\ref{lem:passBar}, so we omit them in favour of a brief overview.

We begin by computing on $\hat{w}_0$. During Stages $\dcone$ and $\dctwo$ of iteration $k$, we compute $\Vert \hat{w}_k,\oes{a}{{}}{{}} \Vert = \Vert \hat{w}_{k-1},\oes{a}{{}}{{}} \Vert /2$,
by marking every second pair of $\oes{a}{{}}{{}}$ symbols (here we mark the even numbered pairs).  Then in Stages $\dcthree$ and $\dcfour$ we compute 
\begin{equation}\label{eq:doubleCounterRecurrence}
\Vert \hat{w}_k,\oes{x}{{}}{{}} \Vert = \left\lfloor \frac{\Vert \hat{w}_{k-1},\oes{x}{{}}{{}} \Vert}{2} \right\rfloor
\end{equation} 
by marking every second pair of $\oes{x}{{}}{{}}$ symbols (here we mark the odd numbered pairs). We then return to Stage $\dcone$ and iterate until $\Vert\hat{w}_k,\oes{a}{{}}{{}}\Vert = 1$: 
the counter now has an odd value (for the first time) and we detect this in Stage~$\dcthree$. The number of fully completed iterations, and final value for~$k$, is~$k=\log \Vert \hat{w},\oes{a}{{}}{{}} \Vert = {\lceil \log (l+1) \rceil}$. An additional stage restores the parity (by introducing an extra~$\#$ symbol and then deleting it after one round) so that we are reading undotted symbols. Then $\hat{w}_k$ is of the form
\begin{equation*}
\begin{aligned}
\hat{w}_k = & \ 
\osr{1}{{}}{\dcfive}\esr{1}{{}}{\dcfive}\;
\oesm{x}{1}{\dcfive}
\ldots
\oesm{x}{{i}}{\dcfive}\; 
\oes{a}{{}}{\dcfive}\; 
\oesm{a}{{}}{\dcfive}%
\ldots
\oesm{a}{{}}{\dcfive}\; 
\oesm{x}{{i+1}}{\dcfive}%
\ldots
\oesm{x}{{l}}{\dcfive}\;
\oesm{x}{{l+1}}{\dcfive}
\ldots
\oesm{x}{{l+c}}{\dcfive}
\end{aligned}
\end{equation*}

At this point the number $\Vert \hat{w_k},\oes{x}{{}}{{}} \Vert$ of unmarked $\oes{x}{{}}{{}}$ pairs satisfies $0\leqslant \Vert \hat{w_k},\oes{x}{{}}{{}} \Vert \leqslant 1$. To see this, note that $\Vert\hat{w}_0,\oes{a}{{}}{\,}\Vert=2^{\lceil\log (l+1)\rceil}$ and $c \leqslant 2^{\lceil\log (l+1)\rceil}$. Solving Equation~\eqref{eq:doubleCounterRecurrence} for $k=\lceil \log (l+1) \rceil$ gives~0 iff 
$$0 \leqslant 
\Vert \hat{w}_0,\oes{x}{{}}{{}} \Vert = l + c
< 2^{\lceil \log (l+1) \rceil}$$ 
and~1 iff 
$$2^{\lceil \log (l+1) \rceil} \leqslant \Vert \hat{w}_0,\oes{x}{{}}{{}} \Vert  = l + c < 2^{\lceil \log (l+c+1) \rceil} $$%
There are no other possible values for $\Vert \hat{w}_0,\oes{x}{{}}{{}} \Vert$ thus we have only to check whether $\Vert \hat{w_k},\oes{x}{{}}{{}} \Vert$ is 0 or 1. To prepare, in two consecutive rounds we apply the rules
\begin{eqnarray*}
\{
\dwrule{\osr{1}{{}}{\dcfive}}{\osr{1}{{}}{\dcsix}\esr{1}{{}}{\dcsix}},
\dwrule{\os{x}{{}}{\dcfive}}{\threesymb{x}{{}}{\dcsix}},
\dwrule{\osm{x}{{}}{\dcfive}}{\oesm{x}{{}}{\dcsix}},
\dwrule{\os{a}{{}}{\dcfive}}{\oes{a}{{}}{\dcsix}},
\dwrule{\osm{a}{{}}{\dcfive}}{\oes{a}{{}}{\dcsix}}
\}
\end{eqnarray*}
and
\begin{align*} %
 \{
& \dwrule{\osr{1}{{}}{\dcsix}}{\osr{1}{{}}{\dcseven}\esr{1}{{}}{\dcseven}},
\dwrule{\os{x}{{}}{\dcsix}}{\oes{x}{{}}{\dcseven}},
\dwrule{\thirdsymbol{x}{{}}{\dcsix}}{\epsilon},
\dwrule{\osm{x}{{}}{\dcsix}}{\oes{x}{{}}{\dcseven}},
\dwrule{\esm{x}{{}}{\dcsix}}{\oes{x}{{}}{\dcseven}},\\
& \dwrule{\os{a}{{}}{\dcsix}}{\oes{a}{{}}{\dcseven}},
\dwrule{\es{a}{{}}{\dcsix}}{\oes{a}{{}}{\dcseven}},
\dwrule{\osm{a}{{}}{\dcsix}}{\oes{a}{{}}{\dcseven}},
\dwrule{\esm{a}{{}}{\dcsix}}{\oes{a}{{}}{\dcseven}}
\}
\end{align*}%
These rules have the effect of shifting the parity of the read head to dotted symbols iff $\Vert \hat{w_k},\oes{x}{{}}{{}} \Vert = 1$. In addition these rules unmark all marked symbols, as the marks are not needed below.

We have two cases:\\
Case 1: If $\Vert \hat{w_k},\oes{x}{{}}{{}} \Vert = 0$ we do not need to change the counter value in order to satisfy Equation~\eqref{eq:counterValue}. In this case we detect $\Vert \hat{w_k},\oes{x}{{}}{{}} \Vert = 0$ by reading the undotted, barred symbol $\osr{1}{{}}{\dcseven}$. To complete the computation we apply the rules
\begin{equation*}
\{
\dwrule{\osr{1}{{}}{\dcseven}}{\osr{1}{{}}{\dcend}\esr{1}{{}}{\dcend}},
\dwrule{\os{x}{{}}{\dcseven}}{\oes{x}{{}}{\dcend}},
\dwrule{\os{a}{{}}{\dcseven}}{\oes{a}{{}}{\dcend}}
\}
\end{equation*}
Case 2: If $\Vert \hat{w_k},\oes{x}{{}}{{}} \Vert = 1$ we double the counter to satisfy Equation~\eqref{eq:counterValue}. In this case we detect $\Vert \hat{w_k},\oes{x}{{}}{{}} \Vert = 1$ by reading the dotted, barred symbol~$\esr{1}{{}}{\dcseven}$. We then restore odd parity and double the counter value, by applying the rules
\begin{eqnarray*}
\{
\dwrule{\esr{1}{{}}{\dcseven}}{\#\osr{1}{{}}{\dcend}\esr{1}{{}}{\dcend}},
\dwrule{\es{x}{{}}{\dcseven}}{\oes{x}{{}}{\dcend}},
\dwrule{\es{a}{{}}{\dcseven}}{\oes{a}{{}}{\dcend}\;\oes{a}{{}}{\dcend}}\}
\end{eqnarray*}
\qed
\end{proof}

\subsection{Proof of main result}
\begin{theorem}\label{thm:2tagSimCycTag}
Given a cyclic tag system~$C$ that computes in time $t(n)$ on input of length~$n$, where~$n$ is at least the length of~$C$'s longest appendant, then there is a 2-tag system~$T_C$ that simulates the computation of~$C$ and computes in time~$O(t^{2}(n)\log t(n))$.
\end{theorem}
\begin{proof} 
As stated, it is assumed that~$C$'s input length $n$ is at least that of its longest appendant. As part of the 2-tag input encoding in Definition~\ref{def:2tagInputEnc}, the constant number of shorter inputs are assumed to be padded to this length.

Recall, from the beginning of Section~\ref{sec:sim}, the decomposition of a single cyclic tag computation step into the conceptual steps~(i),~(ii) and~(iii). 

Lemma~\ref{lem:simAppend} provides the algorithm for step (i) for the case that $\osr{x}{0}{{}}=\osr{1}{{}}{{}}$. For the other case of $\osr{x}{0}{{}}=\osr{0}{{}}{{}}$ we skip (i). Deciding between the two cases is easily implemented by setting the parity to even iff $\osr{x}{0}{{}}=\osr{1}{{}}{{}}$.

Lemma~\ref{lem:passBar} provides the main mechanics for step (ii). After applying Lemma~\ref{lem:passBar} the barred pair $\osr{x}{{}}{{}}\esr{x}{{}}{{}}$ is either on the left of the counter ($i>0$ in the lemma statement), or else has jumped over the counter and is now on the right ($i=0$).
Lemma~\ref{lem:simAppend} assumes the head is on the left. Therefore if the head is on the right we want to move it to the left so that we can apply Lemma~\ref{lem:simAppend} in subsequent iterations. Therefore as a second part of step (ii) we do the following. In one round we read the entire dataword~$\hat{w}$, and reproduce~$\hat{w}$ exactly except that the underindex is incremented and the length of the barred subword is increased from~2 to~3 by a rule of the form 
$\dwrule{\osr{x}{{}}{{}}}{\osr{x}{{}}{{}}\esr{x}{{}}{{}}\thirdsr{x}{{}}{{}}}$. On the one hand, if in the next round we read dotted counter symbols ($\es{a}{{}}{{}}$) then the head is to the left of the counter; then in one more round we set the parity to odd and we're done. On the other hand, if we read undotted counter symbols ($\os{a}{{}}{{}}$) then we set the parity to odd in one round and furthermore we read the counter an extra time to place it to the right of the head.

Lemata~\ref{lem:passBar} and~\ref{lem:simAppend} state that the counter value should be the next power of two greater than $C$'s dataword length $l$. It is not difficult to check that their proofs also hold for the more general case where the counter value is any power of two greater than  $l$. Therefore these proofs are still applicable if $C$'s dataword drastically decreases in length. In this case the big-Oh time bounds in these proofs depend on the counter value rather than $l$. But since the counter value is~$O(t(n))$ this does not affect our overall complexity analysis.

For step (iii) we introduce a new decoration for 2-tag symbols.
So far, the number $q$ of distinct 2-tag symbols that we have used is dependent on our algorithm. We now increase this number to $pq$ where, as usual, $p$ is the number of appendants of $C$. We create a new symbol set by decorating each 2-tag symbol 
$y \in \{
\osr{x}{{}}{{}},
\esr{x}{{}}{{}},
\os{x}{{}}{{}},
\es{x}{{}}{{}},
\thirdsymbol{x}{{}}{{}},
\osm{x}{{}}{{}},
\esm{x}{{}}{{}},
\os{a}{{}}{{}},
\es{a}{{}}{{}},
\thirdsymbol{a}{{}}{{}},
\osm{a}{{}}{{}},
\esm{a}{{}}{{}},
\# 
\}$ with an integer ${\arbApp}$ for all $0 \leqslant {\arbApp} < p$. Using this, our encoding of an arbitrary cyclic tag system configuration is of the form
\begin{equation*}%
\osra{x}{0}{j}{{\arbApp}}
\esra{x}{0}{j}{{\arbApp}}\; 
\osa{x}{1}{j}{{\arbApp}}
\esa{x}{1}{j}{{\arbApp}}
\ldots
\osa{x}{{i}}{j}{{\arbApp}}
\esa{x}{{i}}{j}{{\arbApp}}\; 
\osa{a}{{}}{j}{{\arbApp}}
\esa{a}{{}}{j}{{\arbApp}}\; 
\osa{a}{{}}{j}{{\arbApp}}
\esa{a}{{}}{j}{{\arbApp}}
\ldots
\osa{a}{{}}{j}{{\arbApp}}
\esa{a}{{}}{j}{{\arbApp}}\; 
\osa{x}{{i+1}}{j}{{\arbApp}}
\esa{x}{{i+1}}{j}{{\arbApp}} %
\ldots
\osa{x}{{l}}{j}{{\arbApp}}
\esa{x}{{l}}{j}{{\arbApp}}
\vspace{1.5ex}\end{equation*}

Steps (i) and (ii) are simulated, while ignoring the value of ${\arbApp}$. (Note that our 2-tag algorithms are easily concatenated by having appropriate~$j$ values at the beginning and end of each algorithm.) Then~$j$ is given a value that signals the completion of steps~(i) and~(ii). Then step~(iii) (incrementing the program marker) is simulated by rules of the form
\begin{equation*}
\{
\dwrule{\osra{x}{{}}{j}{\arbApp}}{\osra{x}{{}}{j}{{{\arbApp}'}}\,\esra{x}{{}}{j}{{{\arbApp}'}}},\,
\dwrule{\osa{x}{{}}{j}{\arbApp}}{\osa{x}{{}}{j}{{\arbApp}'}\,\esa{x}{{}}{j}{{\arbApp}'}},\,
\dwrule{\osa{a}{{}}{j}{\arbApp}}{\osa{a}{{}}{j}{{\arbApp}'}\,\esa{a}{{}}{j}{{\arbApp}'}}
\}
\vspace{1.5ex}\end{equation*}
where ${\arbApp}'=({\arbApp}+1)\mod p$.
Applying these rules for a given $k$ takes only one round, or $O(l)$ timesteps.

In the time analysis of the computation of $T_C$ note that for an arbitrary timestep of $C$ we have $l=O(t(n))$. Therefore via Lemmas~\ref{lem:simAppend} and~\ref{lem:passBar}, and the present proof, $T_C$ simulates a {\em single} step of $C$'s computation in time\linebreak $O(t(n) \log t(n))$.\qed
\end{proof}

We get the proof of Theorem~\ref{thm:2-tagSimTM}, our main result, by combining the statements of Theorems~\ref{lem:CycTagSysSimTM} and~\ref{thm:2tagSimCycTag}. If we combine these Theorems directly we get a time bound that is higher than that of Theorem~\ref{thm:2-tagSimTM}. To get our tighter bound a more careful analysis is required where we substitute space bounds for time bounds whenever possible in the analysis.

\section*{Acknowledgements}
Many thanks to Matthew Cook for finding an omission in the proof of Theorem~\ref{thm:2tagSimCycTag}. Both authors are funded by the Irish Research Council for Science, Engineering and Technology.

\end{document}